\documentclass[prl,twocolumn,floatfix,showpacs]{revtex4}
\usepackage{graphicx}
\usepackage{mathrsfs}
\usepackage{amsmath}
\usepackage{amssymb}
\usepackage{amsfonts}
\usepackage{color}
\usepackage[normalem]{ulem}

\makeatletter

\begin{document}

\title{Bond order solid of two-dimensional dipolar
  fermions} \author{S.~G.~ Bhongale$^{1}$, L.~Mathey$^{2,3}$, Shan-Wen
  Tsai$^{4}$, Charles W. Clark$^{3}$, Erhai Zhao$^{1}$}
\affiliation{$^{1}$School of Physics, Astronomy and Computational
  Sciences,
  George Mason University, Fairfax, VA 22030\\
  $^{2}$Zentrum f\"ur Optische Quantentechnologien and Institut f\"ur Laserphysik, Universit\"at Hamburg, 22761 Hamburg, Germany\\
  $^{3}$Joint Quantum Institute, National Institute of Standards and
  Technology \& University of Maryland, Gaithersburg, MD 20899\\
  $^{4}$Department of Physics and Astronomy, University of California,
  Riverside, CA 92521}
\date{\today}

\begin{abstract} Recent experimental realization of dipolar Fermi gases
near or below quantum degeneracy  provides opportunity to engineer Hubbard-like models
  with long range interactions. Motivated by these experiments,  
  we chart out the theoretical phase diagram of
  interacting dipolar fermions on the square lattice at zero temperature
  and half filling. We show that in addition to  $p$-wave superfluid
  and charge density wave order, two new and exotic
  types of bond order emerge generically in dipolar fermion systems.
  These phases feature homogeneous density but periodic modulations of the kinetic
  hopping energy between nearest or next-nearest neighbors. 
  Similar, but
  manifestly different, phases of two-dimensional correlated
  electrons have previously only been hypothesized and termed
  ``density waves of nonzero angular momentum". Our
  results suggest that these phases can be constructed flexibly with
  dipolar fermions, using currently available experimental techniques.   
\end{abstract}
  
\maketitle
Experimental demonstration of Bose-Einstein condensation of
atomic chromium \cite{chromium} and dysprosium \cite{benlev}, both of
which have large magnetic dipole moments, ushers the ultra-cold dipolar gas
to the arena of quantum emulation
\cite{simu,rey}. A gas of the fermionic isotope of 
dysprosium, $^{161}$Dy, has been cooled below quantum degeneracy \cite{priv}. 
A high space-density gas of $^{40}$K$^{87}$Rb, fermionic molecules with electric
dipole moments, has recently been
produced near quantum degeneracy \cite{jun1} and confined in optical lattice \cite{op-la}. 
Such systems are expected to show a rich array
of quantum phases arising from the long-range and anisotropic nature
of dipole-dipole interaction \cite{baranov1,lahaye,fradkin}. This
uniquely distinguishes the dipolar Fermi gas from other Fermi systems,
e.g. the 2D electron gas, the quantum fluid of $^3$He, and Fermi gases
of alkali atoms with short range interactions.
\begin{figure*}
  \includegraphics[width=1\textwidth]{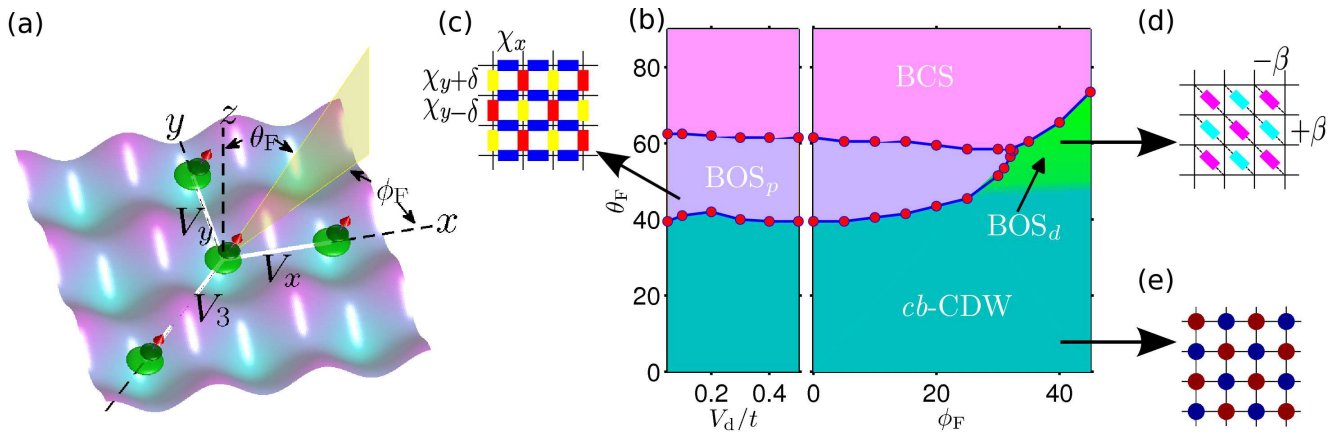}
  \caption{(Color online) Dipolar fermions on square lattice. (a) Schematic of the
    dipolar fermions confined to a square optical lattice
    potential. The induced dipole moment ${\bf d}$ points along the
    direction $\hat{d}=
    \cos\theta_{\text{F}}\hat{z}+\sin\theta_{\text{F}}\cos\phi_{\text{F}}\hat{x}+\sin\theta_{\text{F}}\sin\phi_{\text{F}}\hat{y}$. (b)Phase
    diagram obtained via FRG indicating four phases: $p$-wave bond
    order solid (BOS$_p$), $d$-wave bond order solid (BOS$_{d}$),
    checkerboard charge density wave ($cb$-CDW), and $p$-wave BCS
    superfluid (BCS); left panel-- phase diagram in the
    $\theta_{\text{F}}$-$V_{\text{d}}$ plane at $\phi_{\text{F}}=0$;
    right panel-- phase diagram in the
    $\theta_{\text{F}}$-$\phi_{\text{F}}$ plane at
    $V_{\text{d}}=0.5t$. The phase boundary (solid line) is determined by the 
    abrupt change in the symmetry of the eigenvector of
      the dominant instability (see Fig.~\ref{frg:cdw1}). The smooth crossover from
      $cb$-CDW and BOS$_d$ is indicated by a gradual change of the color shading. (c)-(e)
    Schematic of the bond or density modulation pattern for the
    BOS$_p$, BOS$_{d}$, and $cb$-CDW phase respectively.}
\label{setup}
\end{figure*}
Previous works on dipolar Fermi gases have investigated the
anisotropic Fermi liquid properties
\cite{fradkin,congjun2}, the pairing instability
\cite{baranov2,baranov3,taylor,cooper1,hanpu}, phases showing density
modulation \cite{miyakawa,freericks}, as well as liquid crystal states
\cite{quin,congjun1,erhai}. The possibility of supersolid phases \cite{hofstetter} has
also been discussed.  

For a 2D dipolar Fermi gas on a square lattice at half filling, with
dipole moments perpendicular to the plane, one expects to find a
checkerboard density modulation, known as the charge density wave
(CDW, we follow the nomenclature even though atoms/molecules are
charge neutral). When the dipole moments are aligned in the lattice
plane the system becomes an anisotropic superfluid and the attractive
interaction binds fermions into Cooper pairs. The main question we
address here is, how do different orders compete or cooperate as the
dipole moments are turned from perpendicular to parallel orientation?

We employ the functional renormalization group (FRG) technique
\cite{shankar,zanchi,mathey}, along with self consistent mean field
(SCMF) \cite{ripka} to obtain, for the first time, the
zero-temperature phase diagram of dipolar fermions on a two
dimensional lattice at half filling. The FRG takes an unbiased
approach to treat {\it all} the instabilities of the Fermi surface,
revealing the existence of two new and fascinating quantum phases: the
$p$-wave bond order solid (BOS$_p$); and the $d$-wave bond order
solid (BOS$_{d}$). These bond order solids may be considered as 2D
analogues of the ``bond order wave'' found in the 1D extended Hubbard
model~\cite{nakamura,pinaki,shan-wen1}.

We model single-component dipolar fermions on a two-dimensional square
lattice with lattice constant $a_{\text{L}}$ by the Hamiltonian
\begin{equation}
H=-t\sum_{\langle ij\rangle} a_i^{\dagger}a_j +\frac{1}{2}\sum_{i\ne j}V_{ij}n_in_j,\label{hamiltonian}
\end{equation}
where $t$ represents the nearest neighbor hopping, $a_i$ is the 
fermion annihilation operator at the site $i$, $n_i=a_i^{\dagger}a_i$
is the number operator.  The site index $i$ represents a lattice site
centered at $\mathbf{r}_{i}=i_x a_{\text{L}}\hat{x} + i_ya_{\text{L}}
\hat{y}$, where $i_x$, $i_y$ are integers. The matrix elements of the
dipole interaction in the two-particle Wannier basis are given by
$V_{ij}=\langle ij| V_{\text{dd}}
|ij\rangle=V_{\text{d}}[1-3(\hat{r}_{ij}\cdot
\hat{d})^2]/{(r_{ij}/a_{\text{L}})^3}$, where
$\mathbf{r}_{ij}\equiv\mathbf{r}_{i}-\mathbf{r}_{j}$ and the dipoles
are pointing in the same direction $\hat{d}$. We assume an external
electric or magnetic field ${\bf F}$ pointing in some general
direction.  Then the interaction energy of the dipole moment ${\bf d}$
with the field ${\bf F}$ is equal to $-{\bf F}\cdot{\bf d}$, implying
that the orientation of the dipole moments can be tuned by ${\bf
  F}$. We label the direction of ${\bf d}$ by polar and azimuthal
angles $\theta_{\text{F}}$ and $\phi_{\text{F}}$ respectively, as
illustrated in the schematic of Fig.~\ref{setup}(a).

The interaction between dipoles
can be attractive or repulsive depending on $\theta_{\text{F}}$,
$\phi_{\text{F}}$ and $\mathbf{r}_{ij}$.  For example [refer to
Fig.~\ref{setup}(a)], if $\phi_{\text{F}}=0$, $V_y\equiv
V_{\text{dd}}(a_{\text{L}}\hat{y})$ is always repulsive, while
$V_x\equiv V_{\text{dd}}(a_{\text{L}}\hat{x})$ and $V_3\equiv
V_{\text{dd}} (a_{\text{L}}\hat{x}+a_{\text{L}}\hat{y})$ become
negative for $\theta_{\text{F}}>\vartheta_{\text{c1}}\approx
35.26^\circ$ and
$\theta_{\text{F}}>\vartheta_{\text{c2}}=\cos^{-1}(1/\sqrt{3})\approx
54.74^\circ$ respectively. We shall show that these two critical
points, $\vartheta_{\text{c1}}$ and $\vartheta_{\text{c2}}$, roughly
set the phase boundary between the checkerboard charge density wave
($cb$-CDW), BOS$_p$, and the Bardeen-Cooper-Schrieffer (BCS)
superfluid phase, for the $\phi_{\text{F}}=0$ case.

We now discuss the $T=0$ phase diagram at half filling.  First, we
analyze the weakly interacting limit, $V_{\text{d}} < t$, using FRG.
In this approach, no
assumptions about possible dominant orders are necessary. Rather, the
method includes all processes near the Fermi surface of the
non-interacting system via the generalized 4-point vertex function:
$U_\ell({\bf k}_1,{\bf k}_2,{\bf k}_3)$, where ${\bf k}_{1,2}$ (${\bf
  k}_{3,4}$) are incoming (outgoing) momenta and ${\bf k}_4={\bf
  k}_1+{\bf k}_2-{\bf k}_3$. Here, $\ell$ is the renormalization group
flow parameter that relates the energy cutoff $\Lambda$ to the initial
cutoff $\Lambda_0$ (chosen to be $4t$) via $\Lambda_\ell=\Lambda_0
e^{-\ell}$. Starting with the bare vertex $U_0$, progressively tracing
out the high energy degrees of freedom, a set of coupled
integro-differential equations give the FRG flow for all the vertices.

The renormalized vertex for specific channels of interest, e.g.,
 \begin{equation}
\left.
\begin{array}{l}
  U^{\text{NEST}}_\ell({\bf k}_1,{\bf k}_2)=U_\ell({\bf k}_1,{\bf k}_2,{\bf k}_1+{\bf Q}),\\
  U^{\text{BCS}}_\ell({\bf k}_1,{\bf k}_2)=U_\ell({\bf k}_1,-{\bf k}_1,{\bf k}_2),
\end{array}
\right\}
\label{bcsvertex}
\end{equation}
are extracted by appropriately constraining the in-coming and
out-going momenta. Here ${\bf Q}=(\pi,\pm\pi)$ is the nesting vector
at half filling for the 
square lattice, and $U_{\ell}^{\text{NEST}}$ is the same as
$U_{\ell}^{\text{CDW}}$ of Ref.~\cite{zanchi}.  The channel matrix
with the largest divergent eigenvalue $\lambda$ corresponds to the
most dominant instability of the Fermi liquid. The corresponding
eigenvector $\psi$ defined on the Fermi surface, indicates the
symmetry of the incipient order parameter associated with the
instability.

We perform the FRG analysis for a range of values of $V_{\text{d}}$,
$\theta_{\text{F}}$, and $\phi_{\text{F}}$ producing a 3D phase
diagram, visualized in Fig.~\ref{setup}(b) as slice cuts along two
different planes. To capture and emphasize the key elements of the
phase diagram, first we fix $\phi_{\text{F}}=0$, generating a 2D phase
diagram in the $\theta_{\text{F}}$--$V_{\text{d}}$ plane shown in the
left panel of Fig.~\ref{setup}(b). Next we fix $V_{\text{d}}=0.5t$
instead, yielding the $\theta_{\text{F}}$--$\phi_{\text{F}}$ plane
shown in the right panel of Fig.~\ref{setup}(b).

\begin{figure*}
  \includegraphics[width=1\textwidth]{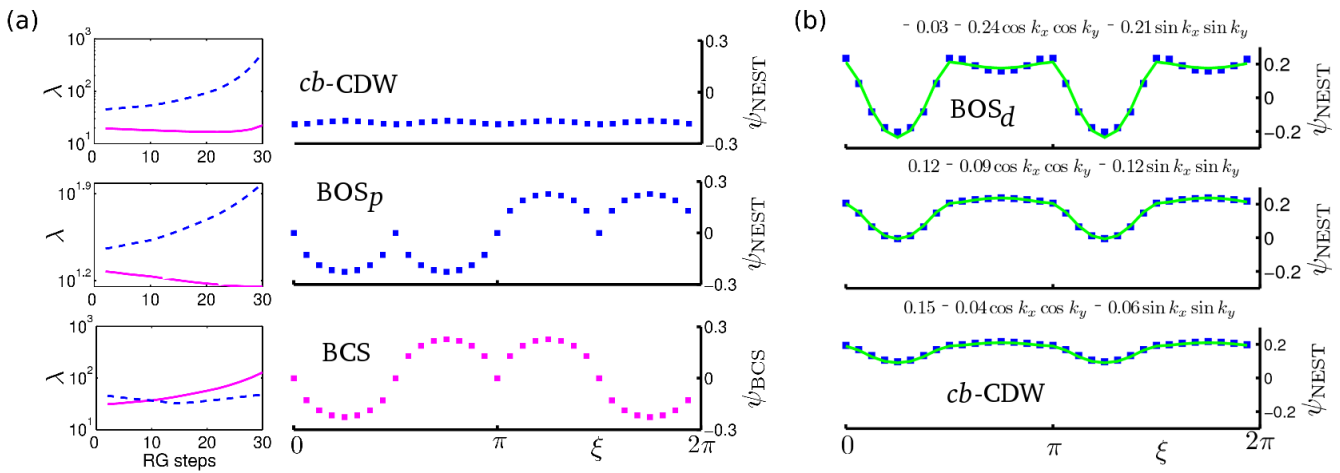}
  \caption{(Color online) FRG results for $V_{\text{d}}=0.5 t$. The FRG is
    implemented numerically by discretizing the Fermi surface into 32
    patches distributed at equally spaced angular points. (a) Top,
    middle and bottom panels represent FRG results for
    $(\theta_{\text{F}},\phi_{\text{F}})=(30^\circ,0),(42^\circ,0)$
    and $(70^\circ,0)$ respectively. Left column: the largest
    eigenvalue $\lambda$ of the $\text{NEST}$ (dashed line) and
    $\text{BCS}$ (solid line) channel. Right column: the corresponding
    eigenvector $\psi$ of the most diverging channel as function of
    $\xi$, the angle of the discrete $\bf{k}$ points on the Fermi
    surface defined by $\tan\xi=k_y/k_x$, plotted with square
    markers. (b) Top, middle and bottom panel represent FRG results
    for
    $(\theta_{\text{F}},\phi_{\text{F}})=(62^\circ,40^{\circ}),(46^\circ,40^{\circ})$
    and $(38^\circ,40^{\circ})$ plotted using square markers. The
    fit is shown in solid line.  As
      $\theta_{\text{F}}$ is increased, $\psi$ smoothly changes from nodeless for
      $\theta_{\text{F}}\lessapprox 46^{\circ}$ to one with nodes for $\theta_{\text{F}}\gtrapprox
      46^{\circ}$.}
\label{frg:cdw1}
\end{figure*}

The $\theta_{\text{F}}$--$V_{\text{d}}$ phase diagram shows the
existence of three phases separated by two critical angles
$\theta_{\text{F}}=\theta_1$ and $\theta_2$, with no appreciable
dependence on $V_{\text{d}}$. For $0\le\theta_{\text{F}}<\theta_1$,
the nesting channel has the largest (most divergent) eigenvalue
$\lambda$. The corresponding eigenvector $\psi_{\text{NEST}}$, as
illustrated in top panel of Fig.~\ref{frg:cdw1}(a), is almost constant
with only small modulation along the Fermi surface.  This implies the
onset of CDW order with $s$-wave symmetry, identified as a
checkerboard modulation of on-site density, the $cb$-CDW shown in Fig.~\ref{setup}(e). The physical
origin of this phase can be traced by observing that
$\theta_1\approx\vartheta_{1\text{c}}$, thus $V_x,V_y,V_3>0$ in this
regime, allowing for a low energy configuration with density
concentrated on the next-to-nearest neighbor sites, consistent with
the perfect nesting of the Fermi surface. For
$\theta_2\le\theta_{\text{F}}\le 90^\circ$, the BCS
channel exhibiting a $p$-wave symmetry is the most diverging under FRG
flow [see Fig.~\ref{frg:cdw1}(a)]. In real space, this corresponds to the
onset of nearest neighbor pairing, $\langle a_i a_{i+\hat{x}}\rangle =
-\langle a_i a_{i-\hat{x}}\rangle$ generated by couplings $V_x$ and
$V_3$, both becoming attractive for $\theta_{\text{F}} >
\theta_{2}\sim\vartheta_{2\text{c}}$. The superfluid phase here is the
lattice analog of the $p$-wave BCS phase discussed previously for
continuum dipolar Fermi gases \cite{baranov1,taylor,hanpu}.

Finally the intermediate regime,
$\theta_1\le\theta_{\text{F}}<\theta_2$, is the most intriguing. The
FRG predicts a leading instability in the nesting channel, similar to
the $cb$-CDW, but instead with a $p$-wave symmetry,
$\psi_{\text{NEST}}({\bf k})\sim \chi ({\bf k})= \chi_0\sin k_y$, as
shown in middle panel of Fig.~\ref{frg:cdw1}(a).  This result suggests
a broken symmetry phase, shown in Fig.~\ref{setup}(c), with periodic
modulation of $\langle a^\dagger_i a_{i+\hat{y}}-\chi_y\rangle =
-\langle a^\dagger_i a_{i-\hat{y}}-\chi_y\rangle= \delta
(-1)^{i_x+i_y}$, where $\chi_y$ is average of $\langle a^\dagger_i
a_{i+\hat{y}}\rangle$ over all bonds. We observe that the nesting
vector ${\bf Q}$ is consistent with the checkerboard pattern of bond
variable representing nearest-neighbor hopping. We refer to this
broken symmetry phase as the $p$-wave bond order solid
(BOS$_p$). Phases with similar, but manifestly different bond order
patterns were conjectured by Nayak and referred to as $p$-density
waves \cite{nayak}.

The right panel of Fig.~\ref{setup}(b),
$\theta_{\text{F}}$--$\phi_{\text{F}}$ phase diagram at fixed
interaction strength, $V_{\text{d}}=0.5t$, shows the three phases
above for small values of $\phi_{\text{F}}$. However, as
$\phi_{\text{F}}$ is increased towards $45^{\circ}$, the BOS$_p$
region shrinks and eventually disappears beyond $\phi_{\text{F}}\sim
35^\circ$. Such change is due to the new features in the dipolar
interactions for $\phi_{\text{F}}$ close to $45^\circ$, where $V_x\sim
V_y$, but the next-to-nearest neighbor interaction along
$\hat{x}+\hat{y}$ and $\hat{x}-\hat{y}$ develop opposite sign. We find
that for such large values of $\phi_{\text{F}}\sim 45^\circ$, the
eigenvector can be fit very well by $\psi_\text{NEST}({\bf k})=\alpha
+ \beta[ \cos k_x\cos k_y+\sin k_x \sin k_y]$, as seen in the right
panel of Fig.~\ref{frg:cdw1}(b). As $\theta_\text{F}$ is increased,
the constant term $\alpha$, which describes the density modulation of
$cb$-CDW order, is gradually reduced, while the magnitude of $\beta$
increases. In the green shaded region in Fig.~\ref{setup}(b), 
$\alpha/\beta$ drops gradually from 1 to 0 as $\theta_F$ is increased
toward the phase boundary to BCS.
We refer to this region where the $\cos k_x \cos k_y$ and $\sin k_x\sin k_y$
components of $\psi_\text{NEST}$ dominant as
the $d$-wave bond order solid (BOS$_{d}$). In this phase, the
density and the nearest hopping $\langle a^\dagger_i
a_{i+\hat{x}/\hat{y}}\rangle$ are homogeneous. But the dipolar
interaction induces an effective diagonal hopping, $\langle
a^\dagger_i a_{i-\hat{x}+\hat{y}}\rangle$, a bond variable with
amplitude proportional to $\beta$ and spatial pattern shown
schematically in Fig.~\ref{setup}(d).  BOS$_{d}$ found here differs
from the $d_{xy}$-density wave conjectured in Ref. \cite{nayak}.

To firmly pin down the nature of the phases, we complement the FRG
analysis with SCMF theory (see Ref. \cite{ripka}) on a square lattice
of finite size $L\times L$ with period boundary condition by defining the normal and pair density
matrices $\rho_{ij}=\langle a_j^{\dagger}a_i\rangle$ and
$m_{ij}=\langle a_ia_j\rangle$ respectively.  The corresponding mean
fields are then given by $\chi_{ji}=-\sum_{kl}\langle jk|
V_{\text{dd}}|li\rangle\rho_{lk}$ and
$\Delta_{ij}=-\frac{1}{2}\sum_{kl}\langle ij| V_{\text{dd}}|kl\rangle
m_{lk}$. The dipole interaction is retained up to a distance of
$10a_{\text{L}}$.
We search for the ground state iteratively by starting with an initial
guess for $\boldsymbol{\rho}$ and $\boldsymbol{m}$, until desired
convergence is reached. The phase boundaries are obtained by
comparing the thermodynamic potential for various converged solutions
(see Supplementary Material).  The chemical potential is tuned to
maintain half filling. And the lattice size $L > 20 a_{\text{L}}$ 
is varied to check the
results do not depend on the choice of $L$.

\begin{figure}[t]
  \includegraphics[scale=.42]{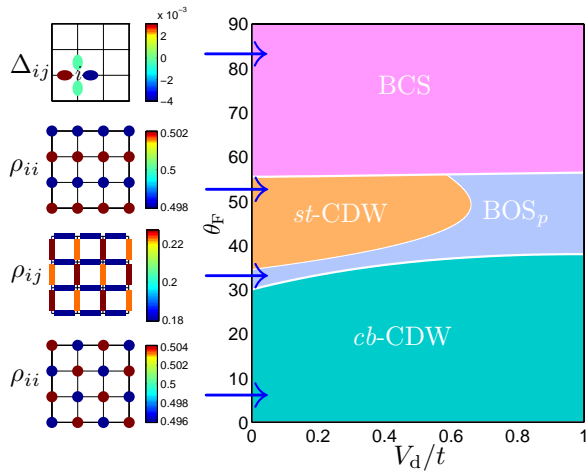}
  \caption{(Color online) SCMF phase diagram. Shown on the left are representatives
    of the on-site density $\rho_{ii}$, the nearest neighbor hopping
    $\rho_{ij}$ (with $j=i+\hat{x}$ or
    $j=i+\hat{y}$), or the pairing gap
    $\Delta_{ij}$
    corresponding to the four phases at $V_{\text{d}}=0.5t$. Lattice size is $32\times 32$. }
\label{mean:phasediagram}
\end{figure}

The SCMF phase diagram for $\phi_{\text{F}}=0$, shown in
Fig.~\ref{mean:phasediagram}, confirms the existence and
interpretation of the three phases found in the FRG
analysis. The phase boundaries are in qualitative agreement
with those from FRG. SCMF for non-zero $\phi_{\text{F}}$
also identifies the BOS$_{d}$ as a phase with the bond modulation
pattern illustrated in Fig.~\ref{setup}(d). We caution
that the SCMF phase diagram is only suggestive.
For example, SCMF predicts an additional striped 
density wave phase, the $st$-CDW, which is not expected
to survive at $V_d\ll t$. This illustrates that SCMF is
insufficient to describe competing orders as opposed to FRG. 
The possibility of $st$-CDW and collapse instability beyond the weak coupling
regime is further discussed in the supplementary material. 

We now provide some intuitive understanding of the bond order phases
by considering a simplified mean field version of
Eq.~(\ref{hamiltonian}), keeping only the nearest neighbor
interactions $V_x$ and $V_y$.  The mean field decoupling of the
interaction term gives $-n_in_j \sim a_i^\dagger a_j a_j^\dagger a_i
\rightarrow \rho_{ij}a_j^\dagger a_i + h.c. - |\rho_{ij}|^2$.  The
modulation of the bond variable, $\rho_{ij}=\langle a^\dagger_i
a_j\rangle$, in the BOS$_p$ phase at $\phi_{\text{F}}=0$ has the form
show in Fig.~\ref{setup}(c), $ \rho_{i,i\pm \hat{x}} = \chi_x$, $
\rho_{i,i\pm \hat{y}} = \chi_y \pm \delta$.  The mean field
Hamiltonian can be written as $H_{R} =-2\sum_\mathbf{k}
\chi_\mathbf{k} b_\mathbf{k}^\dagger a_\mathbf{k} + h.c.$, up to a
constant term.  Here $a_\mathbf{k}$ and $b_\mathbf{k}$ are fermion
annihilation operators defined separately on two sub-lattices related
by the lattice translation vector $a_{\text{L}}\hat{x}$, and
$\chi_\mathbf{k} = (t+V_x\chi_x)\cos k_x +(t+V_y\chi_y)\cos k_y
-iV_y\delta\sin k_y$.  The ground state energy per unit cell is then
given by $E_{\text{GS}} = - 2(\chi_x+\chi_y)(t+V_x+V_y) -
2V_y\delta^2$, clearly indicating that finite bond modulation $\delta$
is energetically favored for positive $V_y$. The
$\phi_{\text{F}}=90^\circ$ situation is identical, only with $x$ and
$y$ axis interchanged, and hence a 90$^\circ$ rotated bond
pattern. Thus, the BOS$_{d}$ phase, with checkerboard pattern of
next-to-nearest bonds near $\phi_{\text{F}}=45^\circ$, naturally
connects the two BOS$_p$ phases on either side.

The bond modulation $\delta$, the energy gap, and the transition
temperature $T_c$ of the BOS$_p$ phase increase with $V_{\text{d}}$
for weak coupling.  Exact diagonalization of Eq. (\ref{hamiltonian})
on a $2\times 8$ and $4\times 4$ cluster with periodic boundary
conditions shows that the optimal place to observe the BOS$_p$ is at
intermediate interaction and tilt angle, e.g. $V_{\text{d}}\sim 2.5 t$
and $(\theta_{\text{F}}, \phi_{\text{F}})=(45^\circ,0^{\circ})$, where
the energy gap, and thus $T_c$, is maximal. Mean field theory estimates
an optimal $T_c \sim 0.23t$, or about $0.05 E_F$ for half filling, 
which is not too far from the temperature achieved in Dy experiment,
$T\sim 0.25 E_F$ \cite{priv}. The BOS$_{d}$ on the other hand is most stable in the
vicinity of $\phi_{\text{F}}=45^{\circ}$ for $\theta_{\text{F}}\sim
60^\circ$.
The characteristic density modulation of the $cb$-CDW and $st$-CDW
phase uniquely distinguishes them from the other phases and may be
detected via in-situ density imaging. The BCS phase can be detected
via pair correlation measurements using noise spectroscopy
\cite{lukin}. Finally the BOS$_{d}$ phase may be distinguished from
BOS$_p$ by probing the $d$-wave symmetry via the pump-probe scheme
discussed in Ref. \cite{demler}. Finally, in the presence of a trap
potential, the insulating plateau at half filling will be surrounded
by metallic regions. The approaches outlined here can be employed to
study dipolar Fermi gas away from half-filling.

SB and EZ are supported by NIST Grant No. 70NANB7H6138 Am 001 and ONR
Grant No. N00014- 09-1-1025A.  LM acknowledges support from the
Landesexzellenzinitiative Hamburg, which is financed by the Science
and Research Foundation Hamburg and supported by the Joachim Herz
Stiftung.  SWT acknowledges support from NSF under grant
DMR-0847801 and from the UC-Lab FRP under award number
09-LR-05-118602.


\begin{thebibliography}{99}
\bibitem{chromium}A.~Griesmaier{\it et al.}, Phys. Rev. Lett. {\bf 94}, 160401 (2005).
\bibitem{benlev}M.~Lu, N.~Q.~Burdick, S.~H.~Youn, and B.~L.~Lev, Phys. Rev. Lett. {\bf 107}, 190401 (2011).
\bibitem{simu}A.~Micheli, G.~K.~Brennen, and P.~Zoller, Nature Physics {\bf 2}, 341 (2006).
\bibitem{rey}A.~V.~Gorshkov {\it et al.}, Phys. Rev. Lett. {\bf 107}, 115301 (2011).
\bibitem{priv} M. Lu, N. Q. Burdick, and B. L. Lev, arXiv:1202.4444.
\bibitem{jun1}K.~-K.~Ni {\it et al.}, Science {\bf 322}, 231 (2008).
\bibitem{op-la} A. Chotia {\it et al.}, arXiv:1110.4420, (2011).
\bibitem{baranov1}M.~A.~Baranov, Phys. Rep. {\bf 464}, 71 (2008).
\bibitem{lahaye}T.~Lahaye {\it et al.}, Rep. Prog. Phys. {\bf 72}, 126401 (2009).
\bibitem{fradkin}B.~M.~Fregoso, and E.~Fradkin, Phys. Rev. Lett. {\bf 103}, 205301 (2009).
\bibitem{congjun2}C.~-K.~Chan, C.~Wu, W.~-C.~Lee, and S.~Das Sarma, Phys. Rev. A {\bf 81}, 023602 (2010). 
\bibitem{baranov2}M.~A.~Baranov, L.~Dobrek, and M.~Lewenstein, Phys. Rev. Lett. {\bf 92}, 250403 (2004).
\bibitem{baranov3}M.~A.~Baranov, M.~S.~Mar'enko, V.~S.~Rychkov, and
  G.~V.~Shlyapnikov, Phys. Rev. A {\bf 66}, 013606 (2002).
\bibitem{taylor}G.~M.~Bruun, and E.~Taylor, Phys. Rev. Lett. {\bf 101}, 245301 (2008).  
\bibitem{cooper1}N.~R.~Cooper, and G.~V.~Shlyapnikov, Phys. Rev. Lett.
{\bf 103}, 155302 (2009).
\bibitem{hanpu}C.~Zhao {\it et al.}, Phys. Rev. A {\bf 81}, 063642 (2010).
\bibitem{miyakawa}Y.~Yamaguchi, T.~Sogo, T.~Ito, and T.~Miyakawa, Phys. Rev. A {\bf 82}, 013643 (2010).
\bibitem{freericks}K.~Mikelsons, and J.~K.~Freericks, Phys. Rev. A {\bf 83}, 043609 (2011).
\bibitem{quin}J.~Quintanilla, S.~T.~Carr, and J.~J.~Betouras, Phys. Rev. A {\bf 79}, 031601(R) (2009).
\bibitem{congjun1}K.~Sun, C.~Wu, and S.~Das Sarma, Phys. Rev. B {\bf 82} 075105 (2010).
\bibitem{erhai}C.~Lin, E.~Zhao, and W. V. Liu, Phys. Rev. B {\bf 81}, 045115 (2010); Phys. Rev. B {\bf 83}, 119901(E) (2011).
\bibitem{hofstetter}L.~He and W.~Hofstetter, Phys. Rev. A {\bf 83}, 053629 (2011).
\bibitem{shankar}R.~Shankar, Rev. Mod. Phys. {\bf 66}, 129 (1994).
\bibitem{zanchi}D.~Zanchi and H.~J.~Schulz, Phys. Rev. B {\bf 61},13609 (2000).
\bibitem{mathey}L.~Mathey, S.~-W.~Tsai, and  A.~H.~Castro Neto, Phys. Rev. Lett. {\bf 97}, 030601 (2006); Phys. Rev. B {\bf 75}, 174516 (2007).
\bibitem{ripka}J.-P. Blaizot and G.~Ripka, {\sl Quantum Theory of
    Finite Systems}, MIT Press, Cambridge MA (1985). 
\bibitem{nakamura}M.~Nakamura, Phys.
Rev. B {\bf 61}, 16377 (2000).
\bibitem{pinaki}P.~Sengupta, A.~W.~Sandvik, and D.~K.~Campbell, Phys. Rev. B {\bf 65}, 155113 (2002).
\bibitem{shan-wen1}K.~-M.~Tam, S.~-W.~Tsai, and D.~K.~Campbell, Phys. Rev. Lett. {\bf 96}, 036408 (2006). 
\bibitem{nayak}C.~Nayak, Phys. Rev. B {\bf 62}, 4880 (2000).
\bibitem{lukin}E.~Altman, E.~Demler, and M.~D.~Lukin, Phys. Rev. A {\bf 70}, 013603 (2004).
\bibitem{demler}D.~Pekker, R.~Sensarma, and E.~Demler, arXiv:0906.0931.
\end{thebibliography}
\end{document}


\title{Bond order solid of two-dimensional dipolar
  fermions--\\Supplementary Material} 
 \author{S.~G.~ Bhongale$^{1}$, L.~Mathey$^{2,3}$, Shan-Wen
  Tsai$^{4}$, Charles W. Clark$^{3}$, Erhai Zhao$^{1}$}
\affiliation{$^{1}$School of Physics, Astronomy and Computational
  Sciences,
  George Mason University, Fairfax, VA 22030\\
  $^{2}$Zentrum f\"ur Optische Quantentechnologien and Institut f\"ur Laserphysik, Universit\"at Hamburg, 22761 Hamburg, Germany\\
  $^{3}$Joint Quantum Institute, National Institute of Standards and
  Technology \& University of Maryland, Gaithersburg, MD 20899\\
  $^{4}$Department of Physics and Astronomy, University of California,
  Riverside, CA 92521}
\maketitle
\section{Functional Renormalization Group}

Functional renormalization group (FRG) is a powerful approach to
correlated fermion systems. In recent years, it has been applied
successfully to a wide of range of problems to yield decisive insights
impossible to obtain from other approaches \cite{rev}.  Our
implementation of FRG follows Zanchi and Schulz \cite{zanchi}, which
can be viewed as a generalization of earlier renormalization group
approach to interacting fermions, e.g., by Shankar \cite{shankar} and
Polchinski \cite{kwp}. Our implementation has been employed to study,
for example, the quantum phases of Bose-Fermi mixtures
\cite{ludwigbosefermi}.

Here we provide further technical details of our numerical FRG
calculation.  The Fermi surface of the non-interacting system is a
nested square for half filling.  It is discretized into $N=32$ patches
with momenta ${\bf k}_i$, $i\in\{1,2,..N\}$. The 4-point interaction
vertex as a function of three momenta $U_\ell({\bf k}_1,{\bf k}_2,{\bf
  k}_3)$ is represented as a $N^3\times N^3$ matrix. We start with the
bare vertices defined on the discretized Fermi surface,
$U_{\ell=0}({\bf k}_i,{\bf k}_j,{\bf k}_l)=V_{\text{dd}}({\bf
  k}_i-{\bf k}_l)$, where $V_{\text{dd}}({\bf q})$ is the dipole
interaction in momentum space, i.e., the lattice Fourier transform of
$V_{\text{dd}}({\bf r}_{ij})$ given in the main text, with proper
anti-symmetrization required by Fermi statistics.  The momentum
dependence of the dipolar interaction is fully taken into account.  At
each FRG step $\ell$, the renormalized vertices $U_{\ell}({\bf
  k}_i,{\bf k}_j,{\bf k}_l)$ are calculated at one-loop level,
truncating the effective action at the six fermion terms and
neglecting self-energy correction \cite{zanchi}. The renormalized
vertices for particular channels, e.g. $U_{\ell}^{\text{NEST}}$ and
$U_{\ell}^{\text{BCS}}$, are then extracted by appropriately choosing
the in-coming and out-going momenta. More technical details on the
flow equations can be found in Ref. \cite{zanchi}. We emphasize that
the traditional single-channel (in either particle-particle or
particle-hole channel) RG, or Random Phase Approximation, is
insufficient to reveal complex order such as BOS$_p$ and BOS$_{d}$ as
discussed here.

Within our implementation of FRG, the shape of the Fermi surface is
fixed. Self-energy corrections in principle can distort the Fermi 
surface, potentially making the nesting at $\bf{Q}=(\pi,\pi)$ less perfect
or even favoring other nesting such as $(0,\pi)$.
These are higher order effects that can be safely neglected in the weak coupling limit. 
Similar arguments were made in many FRG studies
e.g. on the Hubbard model \cite{zanchi}. The FRG phase diagram is
most reliable and trustworthy in the weak
coupling regime, $V_d<<t$. The phase diagram is expected to be modified 
at stronger interactions.

\section{Self-consistent Mean field theory}
The self-consistent mean field theory adopted here follows Ripka et al
\cite{ripka}. It allows coexistence of multiple orders (in particular
bond and current order) and long range interactions.  We define the
real space generalized density matrix on a square lattice of size $L
\times L$ with periodic boundary conditions,
\begin{eqnarray}
\mathcal{R}&=&\left(
\begin{array}{cc}\boldsymbol{\rho} &\boldsymbol{m}\\
-\boldsymbol{m}^*& \mathbf{1}-\boldsymbol{\rho}^*
\end{array}
\right).
\end{eqnarray}
The matrices $\boldsymbol{\rho}$ and $\boldsymbol{m}$ are of dimension
$L^2\times L^2$ and defined as $\rho_{ij}\equiv\langle
{a}_j^{\dagger}{a}_i\rangle$ and $m_{ij}\equiv\langle
{a}_i{a}_j\rangle$, and represent normal and anomalous averages,
respectively.  $\mathbf{1}$ is the $L^2\times L^2$ identity matrix.
$\mathcal{R}$ is thus of dimension $(2L^2)\times (2L^2)$.  Note that
$\rho_{ij}$ include both on-site density $\rho_{ii}$ and nearest
neighbor hopping, e.g., $\rho_{i,i+\hat{x}}$.  The mean field
Hamiltonian of Eq. (1) in the main text is
\begin{eqnarray}
 H_{mf}&=&\left(
\begin{array}{cc}
-\boldsymbol{t}+\boldsymbol{\chi}-\mu \mathbf{1} &\boldsymbol{\Delta}\\
-\boldsymbol{\Delta}^*& -(-\boldsymbol{t}+\boldsymbol{\chi}-\mu\mathbf{1})
\end{array}
\right)
\end{eqnarray}
with the mean fields $\Delta_{ji}=-\frac{1}{2}\sum_{kl}V_{jikl}m_{lk}$
and $\chi_{ji}=-\sum_{kl}V_{jkli}\rho_{lk}$.  $\boldsymbol{t}$ is the
tunneling matrix, i.e., $t_{ij}$ has the value $t$ if the sites $i$
and $j$ are connected by a bond.  We search for the ground state
iteratively by starting with an initial guess for $\boldsymbol{\rho}$
and $\boldsymbol{m}$. Each iteration consists of three steps: (i)
construct $H_{mf}$ using the definitions for $\chi$ and $\Delta$
above, (ii) diagonalize $H_{mf}$ to obtain the eigenvectors $W^{\nu}$
with negative energy eigenvalues $E_\nu$, $H_{mf} W^{\nu}
=-|E_{\nu}|W^{\nu}$, and (iii) update $\boldsymbol{\rho}$ and
$\boldsymbol{m}$ by constructing $\mathcal{R}$ via
$\mathcal{R}=\sum_{\nu}W^{\nu}W^{\nu\dagger}$ \cite{ripka}. The
iteration is repeated until desired convergence is reached. The phase
boundary is determined by comparing the thermodynamic potential of
different converged solutions,
$\mathcal{F}=\text{Tr}\left[(-\boldsymbol{t}-\mu
  \mathbf{1}+\boldsymbol{\chi} )\boldsymbol{\rho}-\boldsymbol{\Delta}
  \boldsymbol{m}^*\right]
-\frac{1}{4}\sum_{ijkl}V_{ijkl}(2\rho_{ki}\rho_{lj}-m_{ji}^*m_{lk})$.
The chemical potential is tuned to maintain half filling. And the
lattice size $L$ is varied to check the results do not depend on the
choice of $L$.

We caution that the SCMF phase diagram is only suggestive. It shows
candidate phases that are energetically competitive on the mean field
level.  On one hand, it can take into account the self-energy
corrections (on the Hartree-Fock-Bogoliubov level) which become
increasingly important for larger values of $V_d/t$. It is assuring to
see that BOS$_p$ persist. On the other hand, it is not always
guaranteed to provide accurate predictions especially for intermediate
dipole tilt angles where different orders compete.  For example, it
misleadingly predicts $st$-CDW in the weak coupling limit $V_d\ll
t$. In the striped phase, the on-site density is modulated with period
$2a_L$ in the $y$ direction while remaining translationally invariant
along $x$.  We find its mean-field energy to be very close to the
BOS$_p$ phase.  However, it is important to note $st$-CDW is ruled out
in the weak coupling limit by FRG and on physical grounds that the
Fermi surface deformation required to favor the $(0,\pi)$ nesting
vector becomes negligible at weak coupling.  In other words, we expect
that fluctuations beyond mean field will shrink the region of $st$-CDW
so it cannot survive to $V_d<<t$.

Finally, we comment on the possibilities of $st$-CDW and collapse
instability at finite $V_d/t$, beyond weak coupling. Neither FRG or
SCMF can unambiguously settle these issues. And we have to wait for
future large scale exact diagonalization or quantum Monte Carlo
studies.